\input{aipcheck}

\documentclass[final]{aipproc}

\layoutstyle{6x9}

\usepackage{amssymb}
\begin{document}

\title{Charged Higgs production at the LHC\\ and CP asymmetries}


\author{E. Christova}{
  address={Institute for Nuclear Research and Nuclear Energy, BAS, Sofia 1784, Bulgaria}
}
\author{H. Eberl}{
  address={Institut f\"ur Hochenergiephysik der \"Osterreichischen Akademie der
     Wissenschaften,\\ A-1050 Vienna, Austria}}
\author{\underline{E. Ginina}}{
  address={Institut f\"ur Hochenergiephysik der \"Osterreichischen Akademie der
     Wissenschaften,\\ A-1050 Vienna, Austria}}
\begin{abstract}
We study CP violation in associated production of a charged Higgs boson and a top quark at the LHC,
$pp \to tH^\pm +X$. The asymmetry between the total cross sections for $H^+$ and $H^-$ production
at next-to-leading order in the MSSM is calculated analytically and a detailed numerical analysis
is performed. Furthermore, subsequent decays of $H^\pm$ to $tb$ and $ \tau^\pm \nu_\tau$ are
considered. In the case with $H^\pm \to tb$ decay the asymmetry can go up to $\sim$~12\%.
\end{abstract}

\maketitle


\section{Introduction}
Complex MSSM parameters can induce CP violation (CPV) in charged Higgs production
associated with a top quark at the LHC, $pp \to tH^\pm
+X$,~\cite{Jennifer,jung:lee:song}. At parton level this process, based on bottom-gluon fusion\footnote{For $m_{H^+}\ge 400$ GeV this is the leading parton level process.}, contains similar CPV contributions as
 the decay $H^\pm \to t b$, which was already studied in~\cite{Hplustb}. Therefore, one would expect CPV effects of the same order, but in the production we have additional box graph contributions. First we study CPV in the $H^\pm$ production
 process only. Then we combine it with subsequent decays into $t b$ and $\tau^\pm \nu_\tau$. We present numerical results for the CP asymmetry induced by vertex, selfenergy and box corrections in the MSSM.
\section{The process}
We consider associate production of a charged Higgs boson and a top quark at the LHC,
\begin{eqnarray}
p\, (P_A)+p\, (P_B)\rightarrow t, \bar{t}~(p_t)+H^\pm(p_{H^\pm})+X~, \label{hadronprocesses}
\end{eqnarray}
based on bottom-gluon fusion at parton level
\begin{eqnarray}
b_r ({\bar b}_r) + g_\mu^\alpha \longrightarrow  t_s ({\bar t}_s) + H^\pm \label{pro}\,,
\end{eqnarray}
with the colour indices $r,s=1,2,3$ and $\alpha=1,...,8.$ At tree-level the process (\ref{pro}) contains two graphs: with bottom quark exchange ($s$-channel),
and top quark exchange ($t$-channel).
\section{CP violating asymmetries}
We define the CPV asymmetry in the $H^\pm$ production as the difference between the total number of produced
$H^+$ and $H^-$ in proton-proton collisions:
\begin{eqnarray}
A_P^{CP}={\sigma(pp\rightarrow \bar{t}H^+) -\sigma(pp\rightarrow t
H^-)\over \sigma(pp\rightarrow \bar{t}H^+) +\sigma(pp\rightarrow t
H^-)}\,, \label{hadronasymm}
\end{eqnarray}
where $\sigma(pp\rightarrow \bar{t}H^+)$ and $\sigma(pp\rightarrow t
H^-)$ are the total cross sections for $H^+$ and $H^-$ production at the LHC.
The asymmetry (\ref{hadronasymm}) is caused by corrections to the
$H^\pm tb$-vertex, selfenergy loops on the $H\pm$-line and box diagrams to both $s$- and $t$-channels in the MSSM
with complex parameters~\cite{progresswork}.

Furthermore we define the CPV asymmetry in charged Higgs boson production in $pp \to tH^\pm$ with a subsequent decay
 $H^\pm \to f$, assuming CPV in both production and decay,
\begin{equation}
A^{CP}_{f}={\sigma(pp\rightarrow \bar{t}H^+\to  \bar{t}f)
-\sigma(pp\rightarrow t H^-\to t \bar f)\over
\sigma(pp\rightarrow \bar{t}H^+\to  \bar{t} f)
+\sigma(pp\rightarrow t H^-\to t \bar f)}\,,
\label{panddasymm}
\end{equation}
where $f$ stands for the chosen decay mode: $f=t\bar b;\, \tau^+ \nu_\tau$ and $ W^+ h^0$. In
narrow width approximation the asymmetry (\ref{panddasymm}) is an algebraic sum of the CPV
asymmetry $A^{CP}_P$ in the production,  and the
CPV asymmetry $A^{CP}_{D,f}$ in the decay $f$ of the charged Higgs boson,
\begin{equation}
A^{CP}_f= A_{P}^{CP}+A^{CP}_{D,f}. \label{finalf}
\end{equation}

\section{Numerical results}

The numerical code based on our analytical results is checked using FeynArts and FormCalc~\cite{FeynArts}.
The Yukawa couplings of the third generation
quarks ($h_t$, $h_b$) are taken to be running~\cite{Hplustb}, at the scale $Q =
m_{H^+} + m_t$. For the evaluation of the parton distribution functions (PDF's) we
use CTEQ6L~\cite{PDFs}, with next-to-leading order $\alpha_s$, at the
same scale $Q$. We assume GUT relation between $M_1$ and $M_2$ and take them both real. The
contributions from diagrams with $\tilde \chi^+$, $\tilde \chi^0$, $\tilde \tau$
and $\tilde \nu$ are
negligible and therefore only contributions from diagrams with $\tilde t$, $ \tilde b$
and $\tilde g$ are shown. If not specified otherwise, we fix the following
MSSM parameters: $ M_2=300~{\rm GeV},~M_3=727~{\rm GeV},~M_{\tilde
U}=M_{\tilde Q}=M_{\tilde D}=350~ {\rm GeV},~\mu=-700~{\rm
GeV},~|A_t| = |A_b| =700~{\rm GeV}, \tan \beta=5,~\phi_{A_t}=\pi/2,~\phi_{A_b}=\phi_{\mu}=0$. The relevant masses of the
sparticles for this choice of parameters, $\tan \beta = 5$ or
30 are shown in Table 1 of~\cite{progresswork}. Our numerical results are in agreements with those shown in~\cite{Jennifer}, but we disagree analytically and numerically with the results given in~\cite{jung:lee:song}.

\subsection{Production asymmetry}
\begin{figure}[h!]
  \includegraphics[height=.2\textheight]{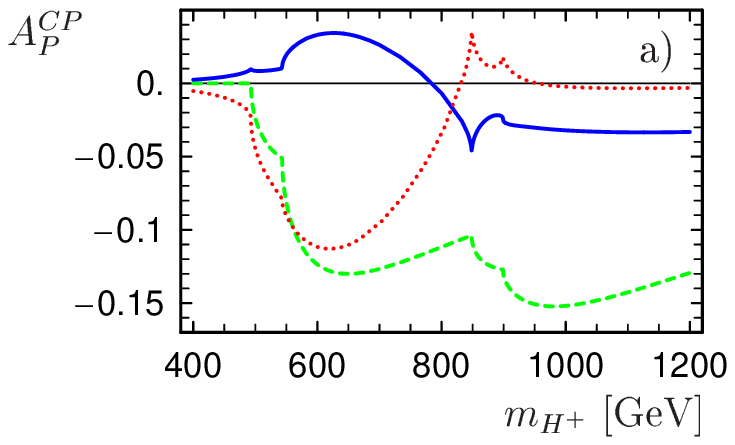}
   \includegraphics[height=.2\textheight]{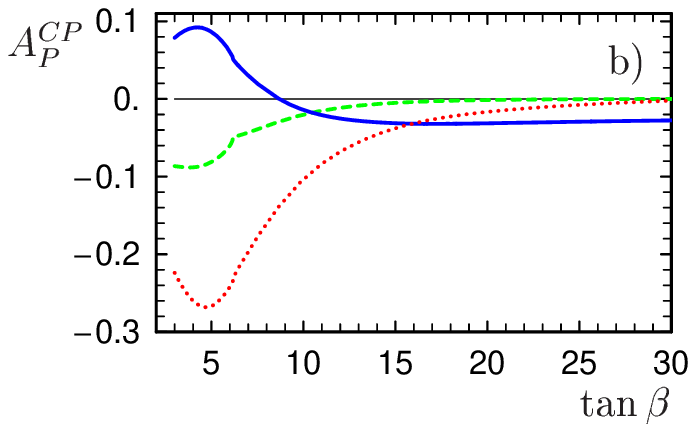}
\caption{The various contributions to the asymmetry $A_P^{CP}$
at hadron level for the chosen set of parameters: ~a)~as a function of $m_{H^+}$; ~b)~ as a function of $\tan\beta$, $m_{\tilde g} =$~450~GeV,
$m_{H^+} =$~550~GeV. The red
dotted line corresponds to box graphs with gluino, the solid blue
one to the vertex graph with gluino, and the green dashed one to
the $W^\pm - H^\pm$ selfenergy graph with $\tilde t \tilde b$ loop.}
\label{fig1}
\end{figure}

The CPV asymmetry $A_P^{CP}$ in the production process can go up to
$\sim 20$\% for relatively small $m_{H^+}$~\cite{Hplustb}. The contributions of the vertex, selfenergy and box graphs with $\tilde t$, $\tilde b $ and $\tilde g$ to $A_P^{CP}$ as functions of $m_{H^+}$ are shown on Fig.~\ref{fig1}a. The large effect seen on
the figure is mainly due to the phase of $A_t$ and the asymmetry reaches its maximum for a maximal phase $\phi_{A_t}=\pi /2$.
The asymmetry is significant for $\tan \beta =5$ and falls down quickly with increasing $\tan \beta$.
This dependence for $m_{H^+}=550$ GeV is shown on Fig.~\ref{fig1}b.
\subsection{Production and decay asymmetry}
\begin{figure}[h!]
  \includegraphics[height=.2\textheight]{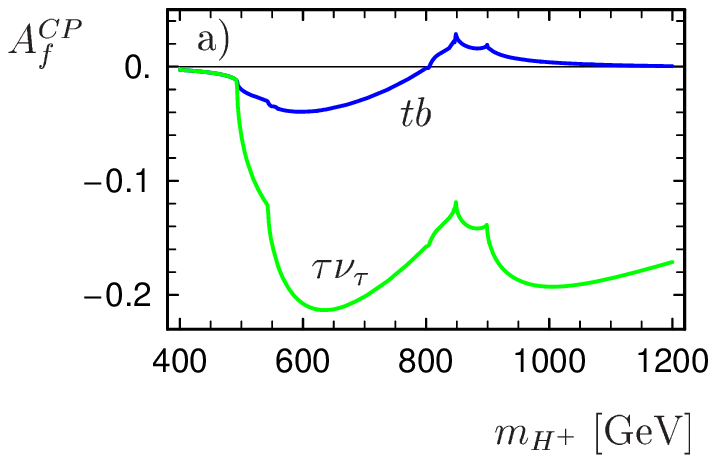}
   \includegraphics[height=.2\textheight]{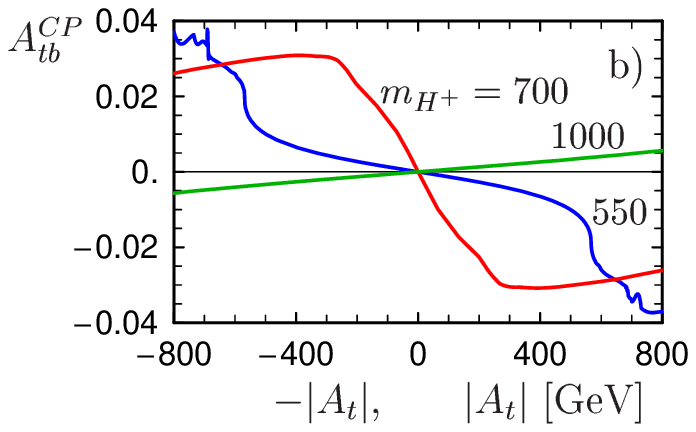}
\caption{The total asymmetry $A^{CP}$ at hadron level for the chosen set of parameters: ~a)~ as a
function of $m_{H^+}$. The blue line corresponds to the case when $H^\pm$ decays to $t b$, and the
green one to $H^\pm$ decay to $ \tau^\pm \nu_\tau $; ~b)~ as a function of $|A_t|$, for three
values of $m_{H^+}$ (in GeV).} \label{fig3}
\end{figure}

The total production and subsequent decay rate asymmetry depends not only on the asymmetry
$A^{CP}$, but also on the branching ratio (BR) of the relevant decay. For small $m_{H^+}$, below
the $\tilde t \tilde b$ threshold, the dominant decay mode is $H^\pm \to t b$, with BR $ \gtrsim
0.9$, while the BR of $H^\pm \to \tau^\pm \nu_\tau$ is in the order of a few percent, decreasing
with increasing $m_{H^+}$. When the $H^\pm \to \tilde t \tilde b$ channels are kinematically
allowed, they start to dominate, and the BR of $H^\pm \to \tau^\pm \nu_\tau$ becomes zero to a good
approximation. However, the BR of  $H^\pm \to t b$ remains stable of the order of 15-20\%, see
Fig.9 in~\cite{progresswork}. In Fig.~\ref{fig3}a we show the total production and decay asymmetry
$A^{CP}_f$ at hadron level, for $f= t b$ and $f=\tau^\pm  \nu_\tau$. The asymmetry $A^{CP}_{
\tau^\pm \nu_\tau}$ can go up to $\sim 20$\% for $m_{H^+}\approx 650$ GeV, but in the shown range
its BR is too small and observation at LHC is impossible.

On the other hand, as the asymmetries in the production only and the in decay in the case of $H^\pm \to tb$ are large and additive~(\ref{finalf}), one would suppose that
the total asymmetry is also large. However, in \cite{progresswork} we show analytically that the $H^\pm - W^\pm$
selfenergy contribution to the asymmetry $A^{CP}_{t b}$
of the decay part cancels exactly the $W^\pm - H^\pm$
selfenergy contribution of the production part. Also the
contributions of the vertex graphs from the production and from the decay can
partially cancel numerically. However, as the box graphs do not have a real analogue in the decay, their contribution remains leading in our studied case.
On Fig.~\ref{fig3}b the dependence of $A^{CP}_{tb}$ on the absolute value of $A_t$ is shown for three different $H^\pm$ masses.

\section{Summary}
We have calculated the CPV asymmetries $A^{CP}_P$, and
$A^{CP}_{f}$, with $f=tb;\,\tau^\pm \nu_\tau$,
between the total cross sections for $H^+$ and $H^-$ production in proton-proton
collisions, proceeding at parton level through $b g$ fusion. We
have performed a detailed numerical analysis, varying different relevant parameters and phases of the MSSM. The asymmetry $A_P^{CP}$ can go up to $\sim 20$\%  at $m_{H^+} \approx  600 ~{\rm GeV}$, $\tan \beta=5$ and a maximal phase of $A_t$. This effect is due to CPV vertex, selfenergy and box contributions with $\tilde t$, $\tilde b$ and $\tilde g$. The total asymmetry in the combined process of production and a subsequent decay is approximately the sum of $A^{CP}_P$ and $A^{CP}_{D, f}$, where $f$ is the relevant decay mode. Despite the fact thast the dominant CPV contribution from the decay cancels with the relevant part of the production,  most promising remains the $tb$ channel. The effect in this case is mainly due to box diagrams with gluino and the asymmetry $A^{CP}_{tb}$ can go up to $\sim 12$\%.

\begin{theacknowledgments}
The authors acknowledge support from EU under the MRTN-CT-2006-035505 network
programme. This work is supported by the "Fonds zur F\"orderung der
wissenschaftlichen Forschung" of Austria, project No. P18959-N16. The work of
E.~C. and E.~G. is partially supported by the
Bulgarian National Science Foundation, grant 288/2008.
\end{theacknowledgments}

\bibliographystyle{aipproc}   

\end{document}